\documentstyle{mn}
\input{psfig}

\def\littleprime{\ifmmode{\scriptscriptstyle \prime }
    \else{\hbox{$\scriptscriptstyle \prime$ }}\fi}
\def\littlecirc{\ifmmode{\scriptscriptstyle \circ }
    \else{\hbox{$\scriptscriptstyle \circ $ }}\fi}
\def\littless{\ifmmode{\scriptscriptstyle s }
    \else{\hbox{$\scriptscriptstyle s $ }}\fi}
\def\arcsec{\raise .9ex \hbox{\littleprime\hskip-3pt\littleprime}}
\def\arcmin{\raise .9ex \hbox{\littleprime}}
\def\degree{\raise .9ex \hbox{\littlecirc}}
\def\arcsecpoint{\hbox to 1pt{}\rlap{\arcsec}.\hbox to 2pt{}}
\def\arcminpoint{\hbox to 1pt{}\rlap{\arcmin}.\hbox to 2pt{}}
\def\degreepoint{\hbox to 1pt{}\rlap{\degree}.\hbox to 2pt{}}
\def\gtapr {\lower .1ex\hbox{\rlap{\raise .6ex\hbox{\hskip .3ex
        {\ifmmode{\scriptscriptstyle >}\else
                {$\scriptscriptstyle >$}\fi}}}
        \kern -.4ex{\ifmmode{\scriptscriptstyle \sim}\else
                {$\scriptscriptstyle\sim$}\fi}}}
\def\ltapr {\lower .1ex\hbox{\rlap{\raise .6ex\hbox{\hskip .3ex
        {\ifmmode{\scriptscriptstyle <}\else    
                {$\scriptscriptstyle <$}\fi}}}
        \kern -.4ex{\ifmmode{\scriptscriptstyle \sim}\else
                {$\scriptscriptstyle\sim$}\fi}}}
\title{The new gravitational lens system B1030+074}
\author[E. Xanthopoulos et al.] {E.~Xanthopoulos,$^{1}$ I.~W.~A.~Browne,$^{1}$
L.~J.~King,$^{1}$ L.~V.~E.~Koopmans,$^{3}$ 
\newauthor 
N.~J.~Jackson,$^{1}$ D.~R.~Marlow,$^{1}$ A.~R.~Patnaik,$^{2}$ R.~W.~Porcas$^{2}$ \& P.~N.~Wilkinson$^{1}$\\
$^{1}$University of Manchester, NRAL Jodrell
Bank, Macclesfield, Cheshire SK11 9DL, England\\ 
$^{2}$ Max-Planck-Institut f\"{u}r Radioastromomie, Auf dem H\"{u}egel 69, D 53121, Bonn, Germany\\
$^{3}$ Kapteyn Astronomical Institute, P. O. Box 800, 9700 AV Groningen, The Netherlands}

\date{Accepted  . 
     Received   }

\begin{document}
\maketitle

\begin{abstract}
We report the discovery of a new double image gravitational lens
system B1030+074 which was found during the Jodrell Bank - VLA  Astrometric Survey (JVAS).
We have collected extensive radio data on the system using the VLA, MERLIN, the EVN and the VLBA and 
optical observations using WFPC2 on the HST. The lensed images are separated 
by 1.56 arcseconds and their flux density ratio  at centimetric wavelengths is approximately 14:1
although the ratio is slightly frequency dependent and the images appear to be time variable.  
The HST pictures show both the lensed images and the lensing galaxy close to the weaker image.
The lensing galaxy has substructure which could be a spiral arm or an interacting galaxy.  

\end{abstract}

\begin{keywords}
gravitation -- galaxies: individual: B1030+074 -- gravitational lensing.
\end{keywords}

\section{Introduction}

The Jodrell-Bank VLA Astrometric Survey (JVAS) is a survey of
flat-spectrum radio sources one of whose purposes is to search for 
gravitational lens systems. JVAS contains the
$\sim$2400 strongest flat spectrum sources in the northern sky
(Patnaik et al. 1992a; Patnaik 1993a; Browne et al. 1998, Wilkinson et al. 1998). 
All sources were observed
with the VLA at 8.4~GHz giving a resolution of 200~mas. Those that were
found to possess multiple components or very complex structure have
been followed-up either with the VLA or with MERLIN and the VLBA to
give high resolution images (King et al., 1998). By this process the
lens systems B0218+257 (Patnaik et al. 1993b), B1422+231 (Patnaik et
al. 1992b) and B1938+666 (King et al. 1997) were discovered. Also in
the survey is the known system MG0414+0534 (Hewitt et al. 1989; Hewitt
et al. 1992) and the probable lens B2114+022 (Augusto et al. 1998). 
The advantage of JVAS compared to the MIT - Green Bank (MG) lens
survey (Bennett et al. 1986) is that by observing only flat spectrum,
hence core-dominated,
sources, it is relatively straightforward to recognize
lens systems, even those which consist of just two images. 
We report here the discovery of such a double system,
B1030+074.  We present VLA, MERLIN and VLBA radio images and EVN
information, together with HST WFPC2 results at 555~nm and 814~nm.  In
Section 2 we present the radio data and the optical data obtained for
B1030+074.  
A discussion of B1030+074 as a lens system follows in Section 3. 

\section{Radio and optical observations}
In this section we present all the radio maps and information obtained
since the discovery of B1030+074.
In Table 1 we present the radio data. Columns 1-4 show the telescope used for various
observations, the date, the frequency and the
resolution, respectively.
The flux density in mJy of the
A and B components as well as the flux density ratio A/B are
shown in columns 5-7.
In Table 2 we present the optical data. Columns 1-4 correspond to the  
same information as in Table 1 while in columns 5-7 we quote the 
Johnson V and Cousins I system magnitudes for the two components and 
their flux density ratio respectively.

\subsection{Radio observations}
All the data were analysed using a combination of 
the NRAO AIPS software package and the Caltech DIFMAP software package 
(Pearson et al. 1994; Shepherd 1997).
The discovery map of B1030+074, made from VLA A-configuration 8.4 GHz
data giving 0.2 arcsec resolution, is shown in Fig. 1.
It shows two distinct compact components with a
separation of 1.56 arcsec, with the fainter image, B, at a PA of 142$^\circ$
relative to image A.
The flux ratio at 8.4 GHz is 12.6.

VLA data for B1030+074 were also obtained  at two more epochs, in 1994 February 22 at 
8.4, 15 and 22 GHz, and 1995 December 20 and 19 at 
15 and 22 GHz respectively. Though both components are detected, each is unresolved in 
the maps. The 15 and 22 GHz maps from 1995  December 20 and 19 are shown in Fig. 1. 
The integrated flux densities were determined using the tasks IMEAN or
JMFIT/IMFIT within AIPS and are also shown in Table 1.

MERLIN data were obtained at 1.7 and 5 GHz in 1993 September 27 and 1996 December 27 respectively
with
resolutions of 150 milliarcsec and 50 milliarcsec respectively.  The
two components of the lens system are again unresolved and a model
fit to the data yields a flux ratio of 18.8 at the 1.7 GHz and
12.0 at the 5 GHz data. The MERLIN 5 GHz map is shown in
Fig. 1.  
\begin{table*}
\caption{
Radio observations of B1030+074.  
 Absolute amplitude errors are estimated to be 5\%; the flux density ratios are accurate to $\approx$1\%. 
}
\begin{tabular}{lcrcccccc}
Telesc. & Obs. date & Frequ.   & Resol. & Flux density of A & Flux density of B & Flux density ratio \\
    &         &  (GHz)   & (arcsec)   &  (mJy)  & (mJy) &    \\
    &         &        &      &      \\
EVN & 1994 05 15  &          1.7 & 0.015 &  147 & 8.1 & 18.1 \\
EVN & 1994 11 18  &          5  &  0.005 & 173 &  10.9 & 15.9 \\
MERLIN & 1993 09 27 & 1.7 & 0.150  &  186 & 9.8 & 18.8    \\
MERLIN & 1996 12 27 &        5  & 0.050 & 326 & 27.3 & 12.0 \\
VLBA & 1995 11 12 &          5   & 0.003 & 248 & 19.1 & 13.0 \\
VLA   & 1992 10 17 &        8.4  & 0.240 & 202 & 16.0 & 12.6 \\
VLA  & 1994 02 22 & 8.4  & 0.240 & 197 & 12.9 & 15.2  \\
VLA  & 1994 02 22 & 15  & 0.140 & 208 & 14.8  & 14.0   \\
VLA  & 1995 12 20 & 15  & 0.140 & 295 & 24.4  & 12.1 \\
VLA  & 1994 02 22 & 22  & 0.080 & 184 & 15.3 & 12.0 \\
VLA  & 1995 12 19 & 22  & 0.080 & 219 & 12.2  & 18.0 \cr 
\end{tabular}
\end{table*}
\begin{table*}
\caption{
Optical observations of B1030+074. 
The transformed Johnson V and Cousins I
magnitudes for the A and B components are presented and the errors are within 0.1 mag.}
\begin{tabular}{lccccccc}
Telesc. & Obs. date & Wavelength & Resol. & Magnitude of A & Magnitude of B & Flux density ratio \\
    &         &  (nm)   & (arcsec)   &   &  &    \\
    &         &        &      &      \\
HST/WFPC2 & 1997 02 03 &  555 & 0.045 & 20.34  & 24.10   & 27.4  \\
HST/WFPC2 & 1997 02 03 &  814 & 0.045 & 18.75 &  22.17   & 23.4  \cr
\end{tabular}
\end{table*}
 
\begin{figure*}
\begin{tabular}{cc}
\psfig{file=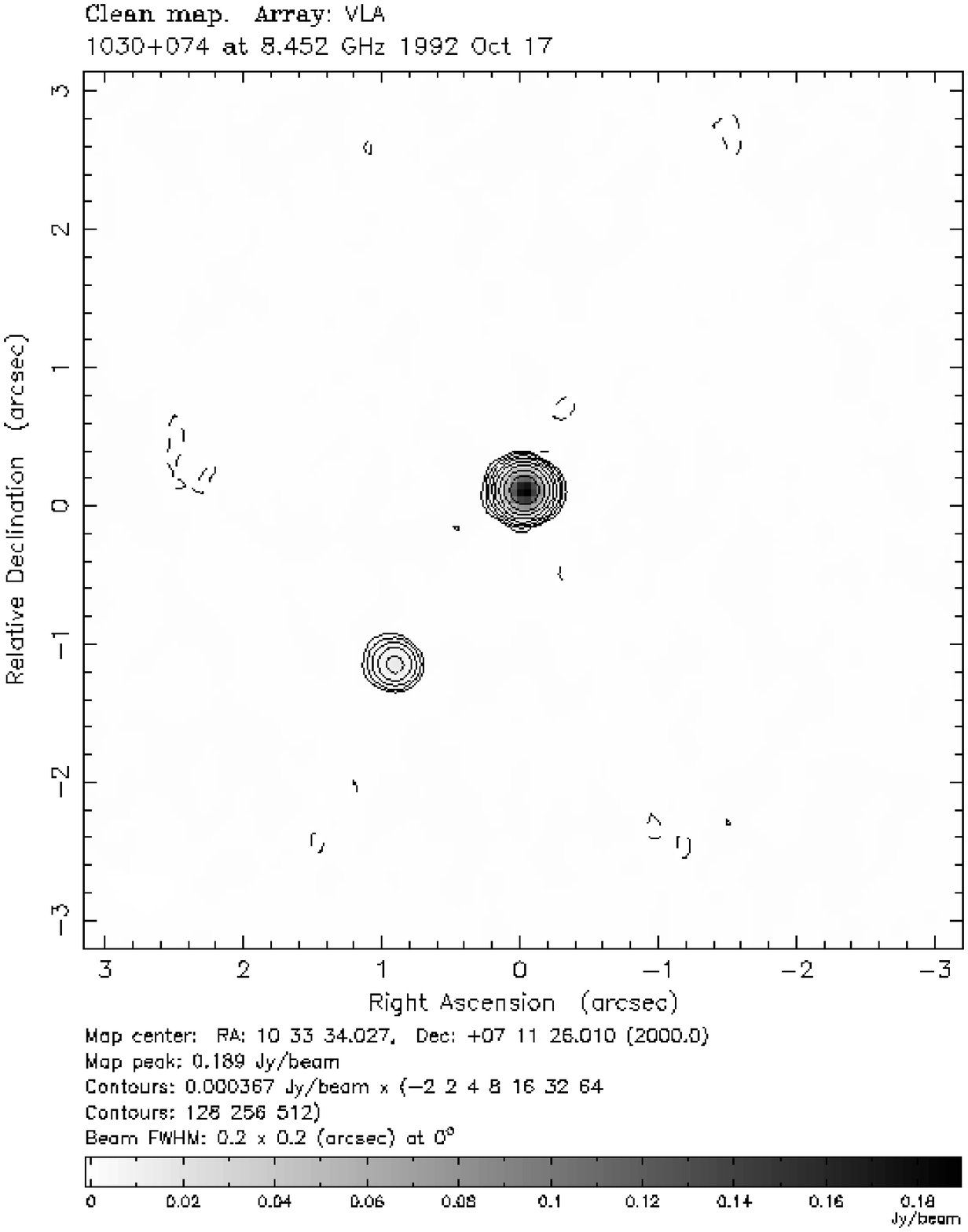,width=9cm,height=9cm,bbllx=0.4in,bblly=2.3in,bburx=8.3in,bbury=9.9in,clip=}&
\psfig{file=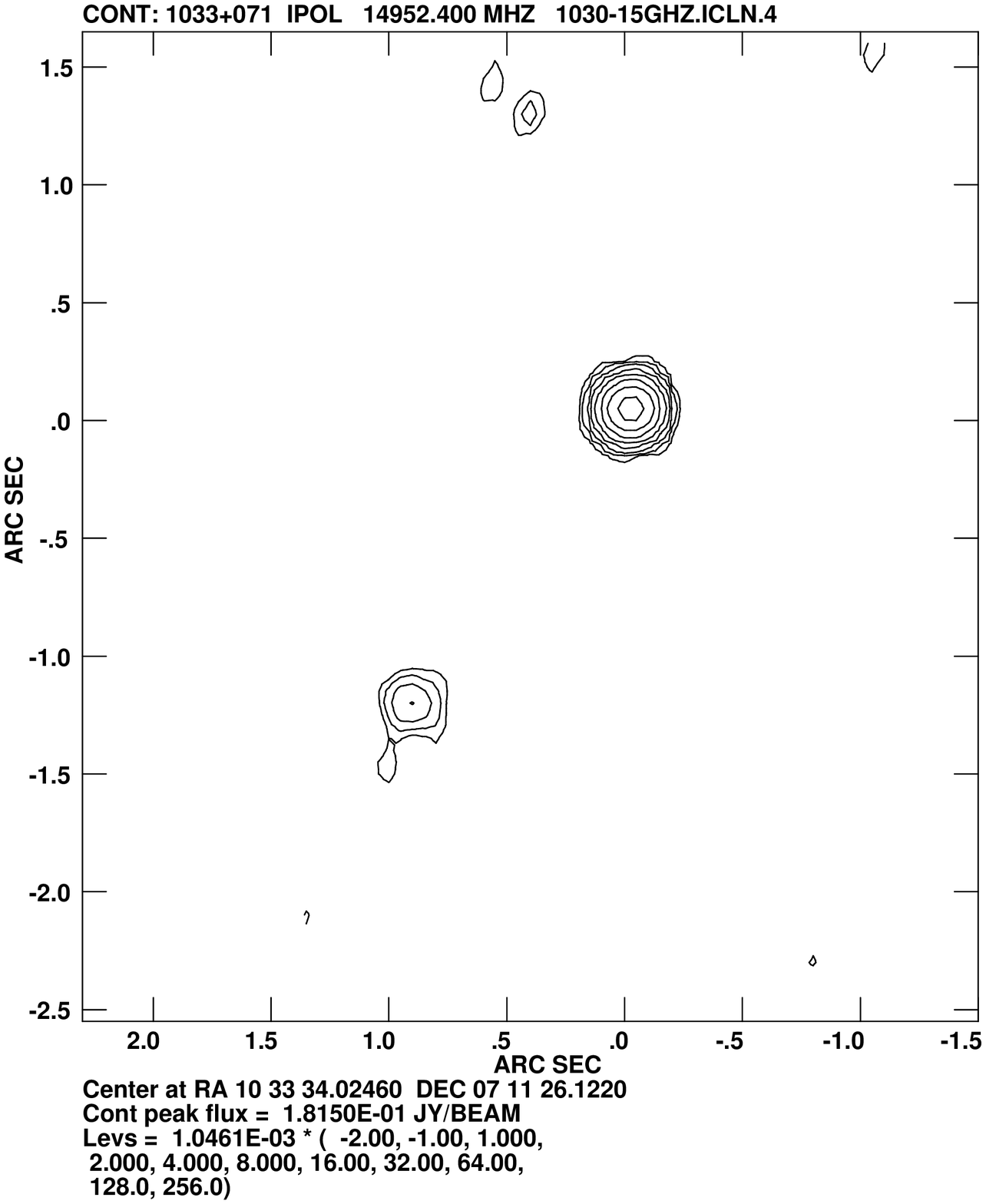,width=9cm,height=9cm,bbllx=0.6in,bblly=1.95in,bburx=8in,bbury=9.8in,clip=}\\
\psfig{file=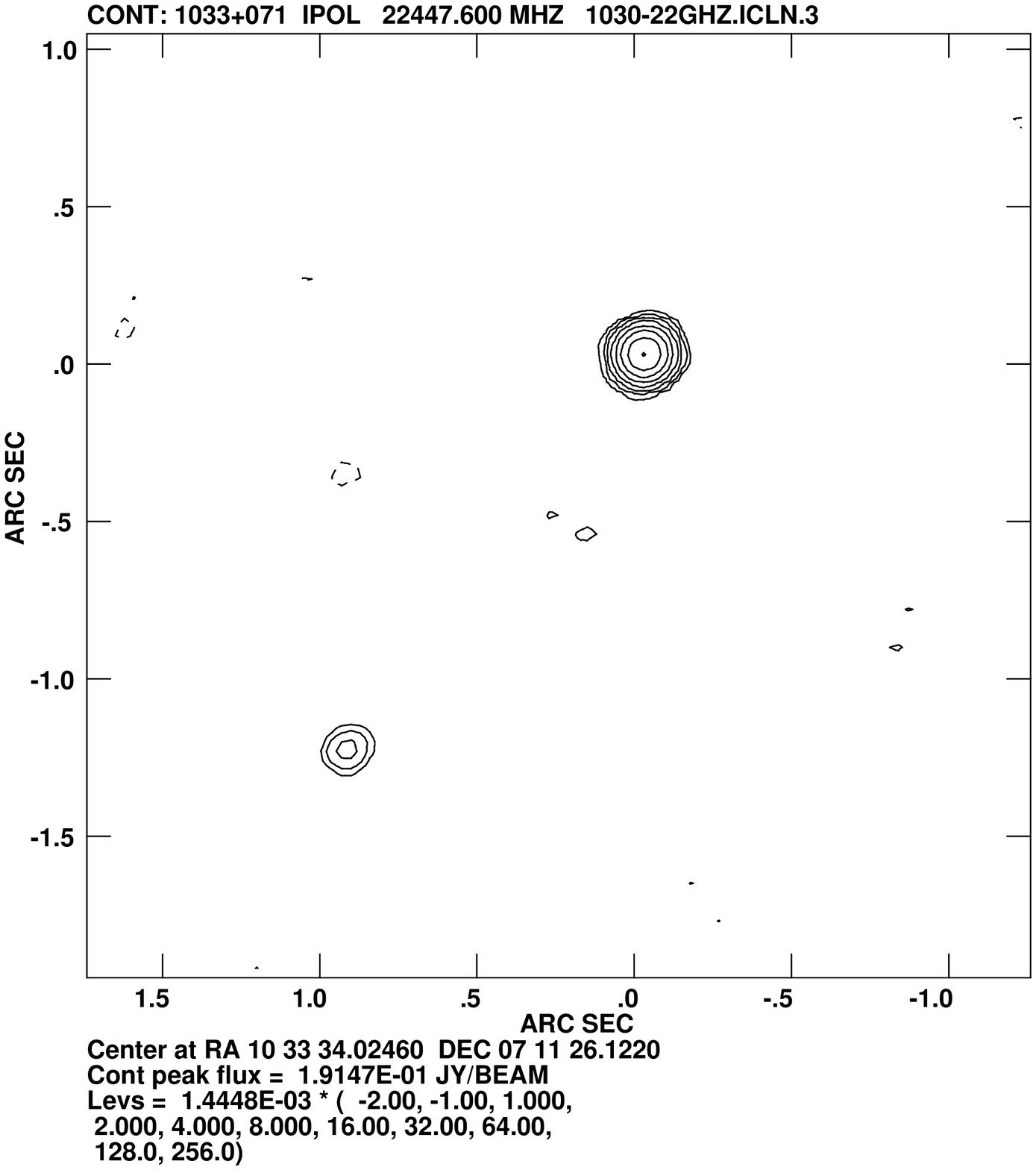,width=9cm,height=9cm,bbllx=0.4in,bblly=2.26in,bburx=8.15in,bbury=9.45in,clip= }&
\psfig{file=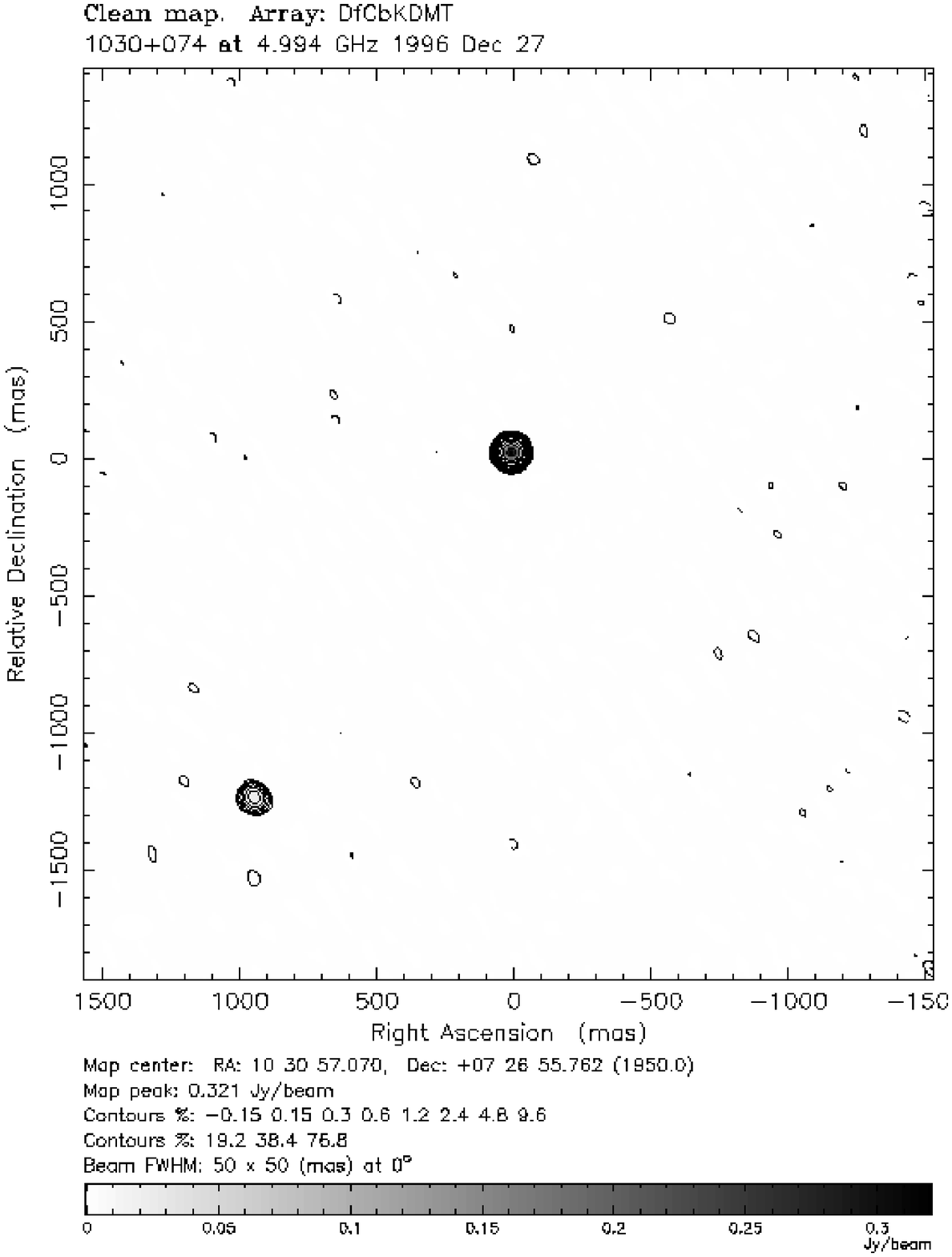,width=9cm,height=9cm,bbllx=0.5in,bblly=2.14in,bburx=8.15in,bbury=10.1in,clip=}\\
\end{tabular}
\caption{The VLA and MERLIN images of the B1030+074 system. 
The J2000 coordinates for B1030+074 are Right
Ascension 10$^{h}$ 33$^{m}$ 34.02460$^{s}$
and
Declination +07\degree 11\arcmin 26.122\arcsec.
Top left: The VLA 8.4 GHz discovery 
map restored with a 200 $\times$ 200 mas beam. The contours are 0.00037 Jy per beam $\times$ (-2, 2, 4, 8, 16, 32, 64, 128, 256, 512), 
and the peak brightness is 0.189 Jy per beam. 
Top right: VLA 15-GHz image. The contours are 0.00105 Jy per beam $\times$ (-2, -1, 1, 2, 4, 8, 16, 32, 64, 128, 256), and 
the peak brightness of the image is 
0.181 Jy per beam (0.140 arcsec resolution). Bottom left: VLA 22-GHz image. The contours are 0.00144 Jy per beam $\times$ (-2, -1, 1, 2, 4, 8, 16, 32, 64, 128, 256), and 
the peak brightness of the image is 0.191 Jy per beam (0.080 arcsec resolution). 
 Bottom right: MERLIN 5-GHz image restored with a 50 $\times$ 50 mas beam. The contours are -0.15, 0.15, 0.3, 0.6, 1.2, 2.4, 4.8, 9.6, 19.2, 38.4 and 76.8 \% of the peak brightness value of 0.321 Jy per beam.}
\end{figure*}

\begin{figure*}
\begin{tabular}{cc}
\psfig{file=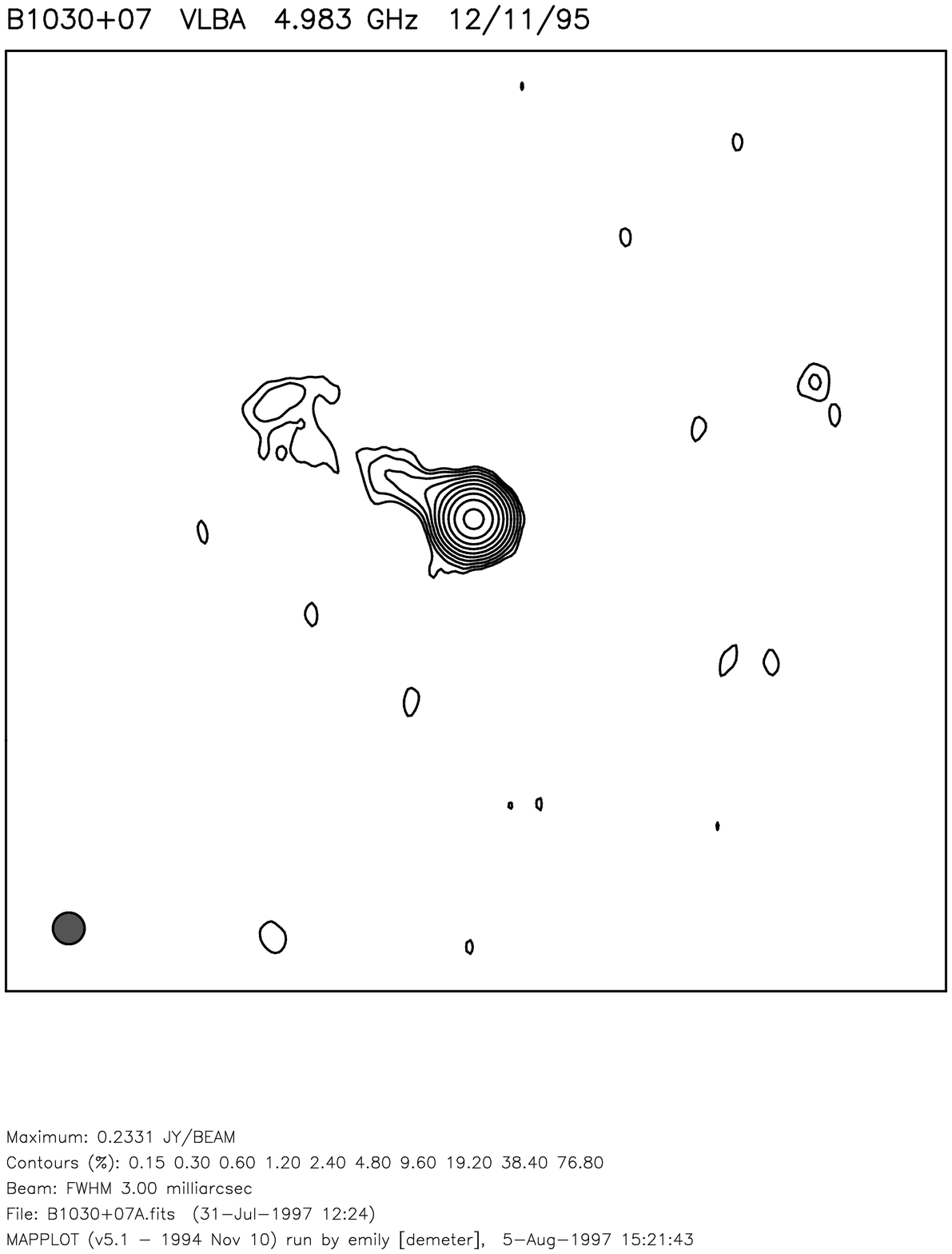,width=9cm,height=9cm,bbllx=1.2in,bblly=2.5in,bburx=8.35in,bbury=9.25in,clip=} &
\psfig{file=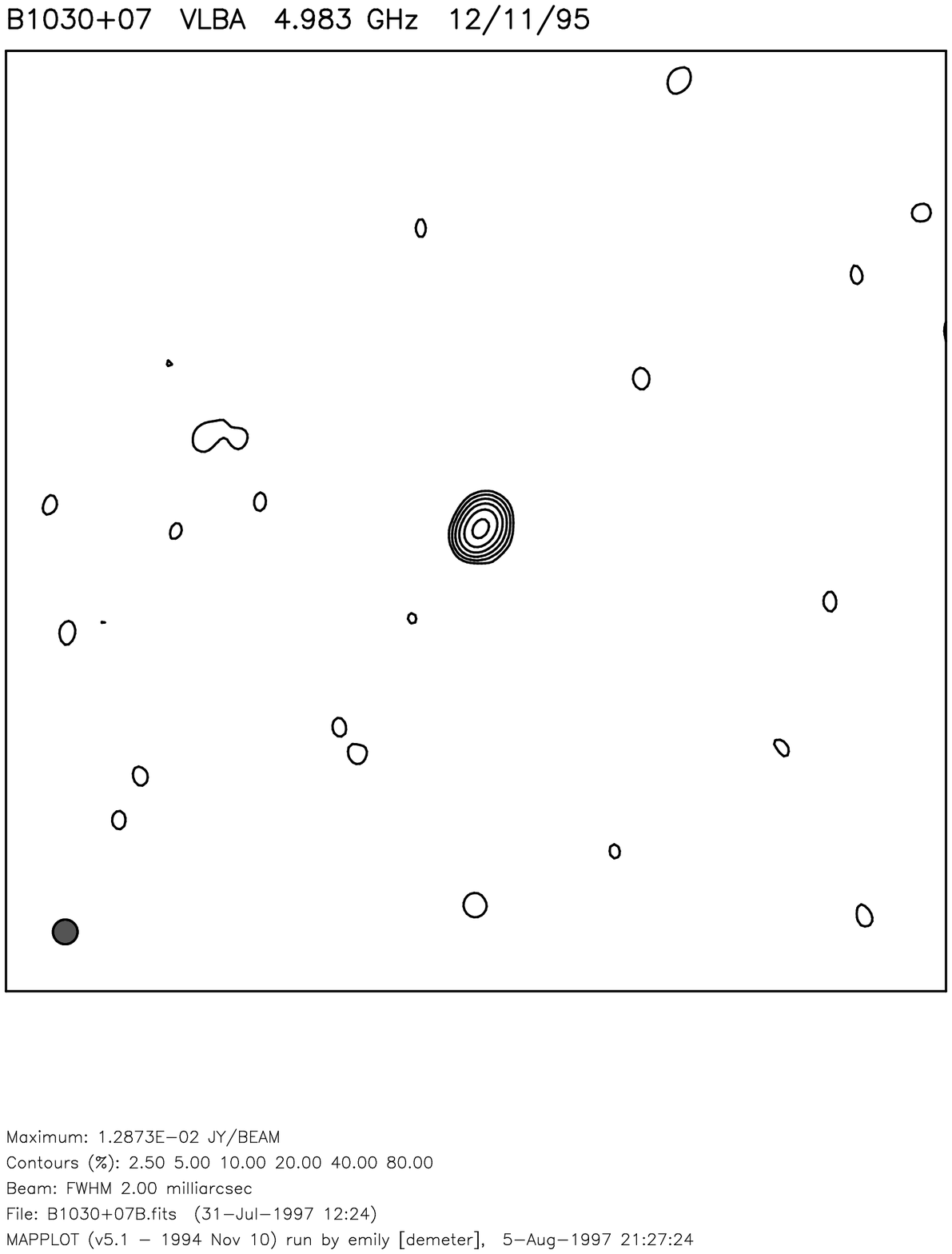,width=9cm,height=9cm,bbllx=1.2in,bblly=2.5in,bburx=8.35in,bbury=9.25in,clip=}\\
\end{tabular}
\caption{The VLBA 5 GHz maps of the two components A and B of the B1030+074 system (resolution 3$\times$3 mas). The contours are set to 0.15, 
0.30, 0.60, 1.20, 2.40, 4.80, 9.60, 19.20, 38.40, 76.80\% of the maximum value of 0.233 Jy per beam for the A component and 2.5, 5, 10, 20, 40,
80\% of the peak brightness of 0.013 Jy per beam for the B component. 
}
\end{figure*}

VLBA observations from 1995 November 12
at 5 GHz with a resolution of 3 mas were able to resolve at least one of 
the components. In
Fig. 2 the stronger A component is seen to have a jet-like
extension to the North-East (PA of $\approx$65\degree) and 20 mas in length.  The
weaker B component remains unresolved
even at this resolution.  From the VLBA data the flux ratio of the two
components is found to be 13.0.

EVN observations at 1.7 with a resolution of 15 mas and 5 GHz with a resolution of 5 mas
were carried out
during 1994 May 15 and November 18 respectively. Both lens components are
detected and are unresolved at 1.7 GHz. The 1.7 GHz data give a
flux density ratio of 18.1 for the two components and a position
separation of 1567$\pm$1 mas (J2000) in position angle 143\degreepoint 4$\pm$0.1.
However, there is a slight extension of the A component visible in the 5 GHz map,
at the same position angle as the VLBA extension.
From the 5 GHz data
we find the flux ratio of the two components to be
15.9 and the position separation is the same as for the 1.7 GHz data.

Since we had VLA data in two epochs at each of the three
frequencies 8.4, 15 and 22 GHz, we were able to look for 
variations in the components. 
There is a definite change in
the flux density ratio and the flux densities of the two components
change with time (see Table 1).   

\subsection{HST observations}
On 1997 February  3, HST images were obtained of B1030+074
using the Wide Field Planetary Camera (WFPC2) in two filters F555W
(nearly Johnson V) and F814W (nearly Cousins I).  Two exposures of 500
seconds each were obtained in each filter. For the reduction of the
data we used STSDAS and other packages within IRAF.\footnote{IRAF is
distributed by the National Optical Astronomy Observatories, which is
operated by the Association of Universities for Research in Astronomy,
Inc. (AURA) under cooperative agreement with the National Science
Foundation.}  The images were rotated, cleared of cosmic ray events
and the two exposures in each filter were averaged. The contour plots of the 
final images
are shown in Fig. 3. 
The lowest contour level was selected to be
2$\sigma$ of the sky background value and the contour levels are
separated by a factor of 2 in intensity.  The HST contour plots of
B1030+074 reveal compact optical objects corresponding to both radio
components, together with a galaxy between them and very close to the B component. These data leave no
doubt that B1030+074 is a gravitational lens system. The lensing
galaxy appears not to have a simple smooth light distribution nor is
it symmetric about the galaxy core. The overall extension of 1.035 arcsec is in a
position angle nearly perpendicular to the image separation; it corresponds to 
6.1 kpc at the redshift of 0.599 of the galaxy (see section 3), 
assuming  H$_{0}$ = 75 km sec$^{-1}$
Mpc$^{-1}$ and  q$_{0}$ = 0.
Greyscale images of the two components and
the galaxy in both the I and V filters are also presented in Fig. 3.
The lowest intensity level is selected to be equal to 1$\sigma$ of the
sky background value and the maximum intensity is selected such that  
most of the detailed structure of both the lens and the lensing galaxy is 
revealed. The transfer function between these two points is linear.

As the lensing galaxy is very close to the weak B lens component it is not easy to
separate the galaxy from the lens component. In order 
to derive accurate photometry of the weak component and of the lensing galaxy, 
we computed and subtracted scaled PSFs from the direct CCD images.
By means of the TinyTim program (Krist 1997) we computed an
oversampled numerical PSF. We then used this optimal PSF in order to
subtract, interactively, using programs within AIPS, the B
component,
until we saw no residuals of this component.  More detail of the
lensing galaxy is then revealed. A contour plot of the galaxy is shown in
Fig. 4. The lowest contour level is 3$\sigma$ of the sky background
value while consecutive contours differ by a factor of 2 in intensity.
Although not an edge-on galaxy the lensing configuration bears a
resemblance to that in the B1600+433 system (Jaunsen \& Hjorth
1997).
The extended structure to the West may be either a spiral arm, or a
second interacting galaxy. Given the present resolution it is
difficult to say which but future
observations with the HST using the Advanced Camera might tell us the answer.
Using the program/command ``ellipse" within the STSDAS
package in IRAF we fitted elliptical isophotes to the galaxy.  The
mean isophotal intensity as a function of the semi-major axis in pixels as well as the 
magnitude (logarithm of the isophote flux) as a function of the semi-major axis in pixels and 
the magnitude (logarithm of the isophote flux) as a function of the the semi-major axis (in pixels) to the 1/4 
are also shown in Fig. 4. The profile information seems to support the
spiral galaxy interpretation since an r$^{1/4}$ law does not fit the surface brightness 
profile of the galaxy.  
Conselice (1997) show that there is a strong 
correlation between morphological asymmetry and Hubble type in the sense that later-type
spirals show an increase in asymmetry. Hence, the asymmetry seen in the lens is
further evidence for the spiral morphology. 

Photometry of the A and B components and of 
the lensing galaxy was performed using the AIPS task 
TVSTAT which gives image
statistics within user-defined apertures.
We then followed the
standard HST photometric procedures (Whitmore 1997) and relied upon
the values of the PHOTFLAM and PHOTZPT keywords (that is the flux of a
source with constant flux per unit wavelength in erg s$^{-1}$
cm$^{-2}$ A$^{-1}$, which produces a count rate of 1 DN per second and
the zeropoint of the instrument respectively) appearing in the header
of the WFPC2 frames in order to convert the flux densities into
standard V and I magnitudes.  The individual magnitudes of the A and B
components in the Johnson V and Cousins I systems were derived from
the F555W and F814W magnitudes.  The measurements for the galaxy
result from the integration of the flux remaining after the removal
of the fitted PSF to the B component.  The integrated I magnitude of
the galaxy is 20.44$\pm$0.1.  The galaxy is not clearly visible in the HST V
image, we only see its compact nucleus plus some faint extended
structure which leads to an estimate of $\approx$22 for the total magnitude. 
The V and I
magnitudes of the A component were found to be 20.34$\pm$0.1 and 18.75$\pm$0.07
respectively while those of the B component are fainter by
$\approx$3.5 mag in both filters. The optical flux ratio of the two components
(A/B is 23.4 in the I and
27.4 in the V)
is much greater than the ratio in the radio which is consistent with extinction due to dust from the lensing galaxy. 
The position separation of the two components calculated from the I HST image is 
1569$\pm$2 mas in a position angle 143.5\degreepoint $\pm$0.1, same as that of the radio within the errors quoted, 
while the separation between 
the strong component and the center of the galaxy is 1369$\pm$4 mas in a position angle 142\degreepoint 50$\pm$0.1 and of the 
weak component and the galaxy 168$\pm$4 mas in a position angle of 156\degreepoint $\pm$0.1.   

\begin{figure*}
\caption{The I and V HST images of the B1030+074
system are shown in the top left and right respectively. North is up and East to the left. 
The lowest contour level is 2$\sigma$ of the sky background value and consecutive contours differ by a factor of 2 in 
intensity.
The lensing galaxy lies closer to the weak component and appears to be extended in the almost East-West
direction. 
We only see the nucleus and a little of the structure of the galaxy of B1030+074
in the V HST image. Greyscale images of
the source and the lens are shown in bottom left and right. The greyscale minimum level is set equal to 1$\sigma$ of the 
sky background value.
}
\begin{tabular}{cc}
\setlength{\unitlength}{1mm}
\begin{picture}(480,480)
\put(0,0){\includegraphics{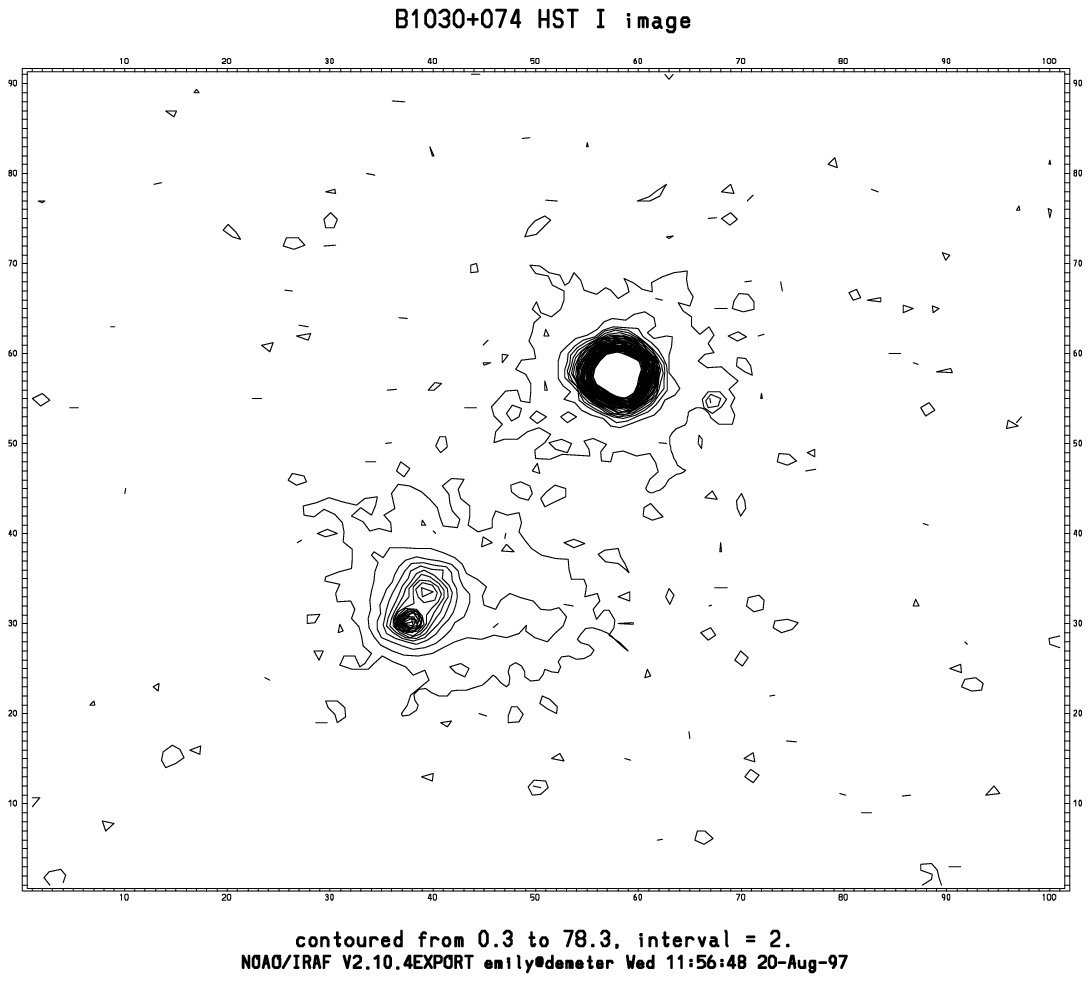}}
\put(100,0){\includegraphics{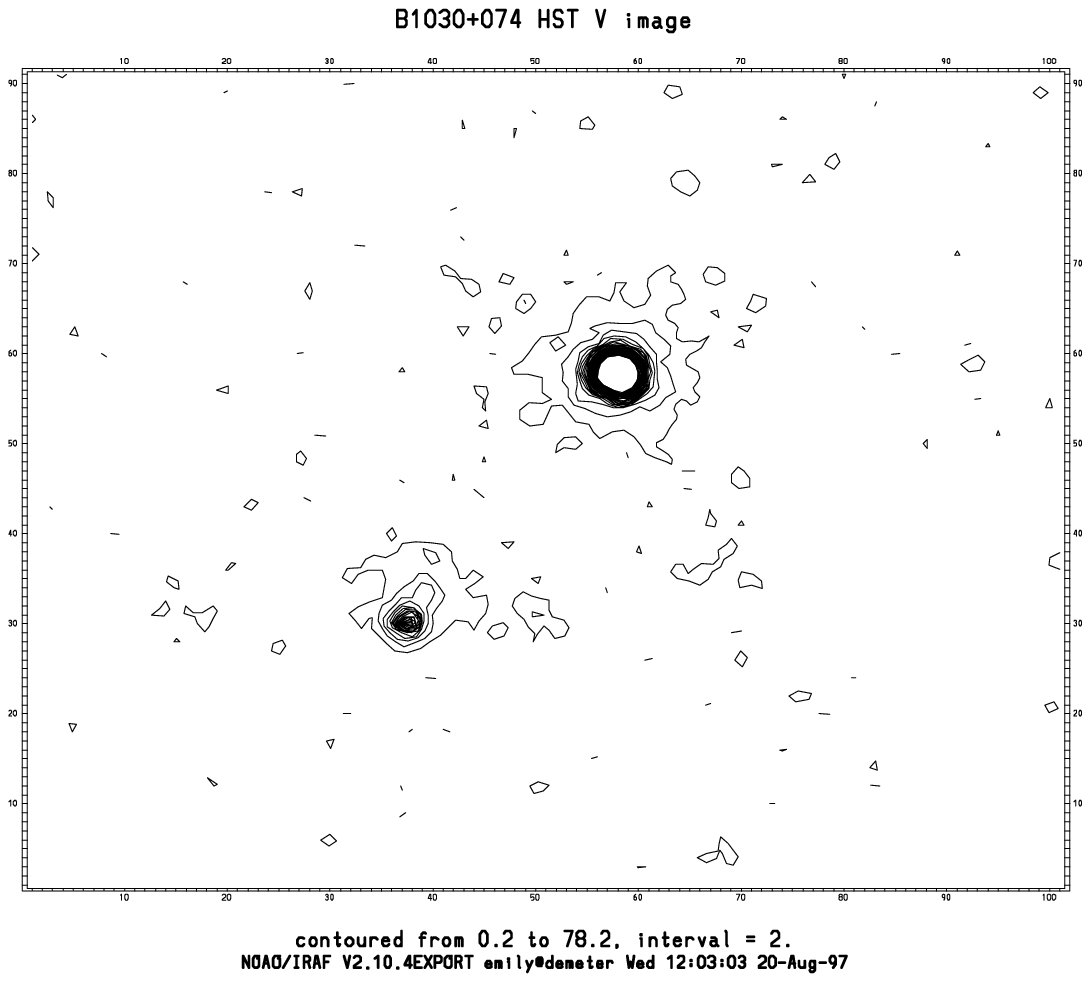}}
\put(0,100){\includegraphics{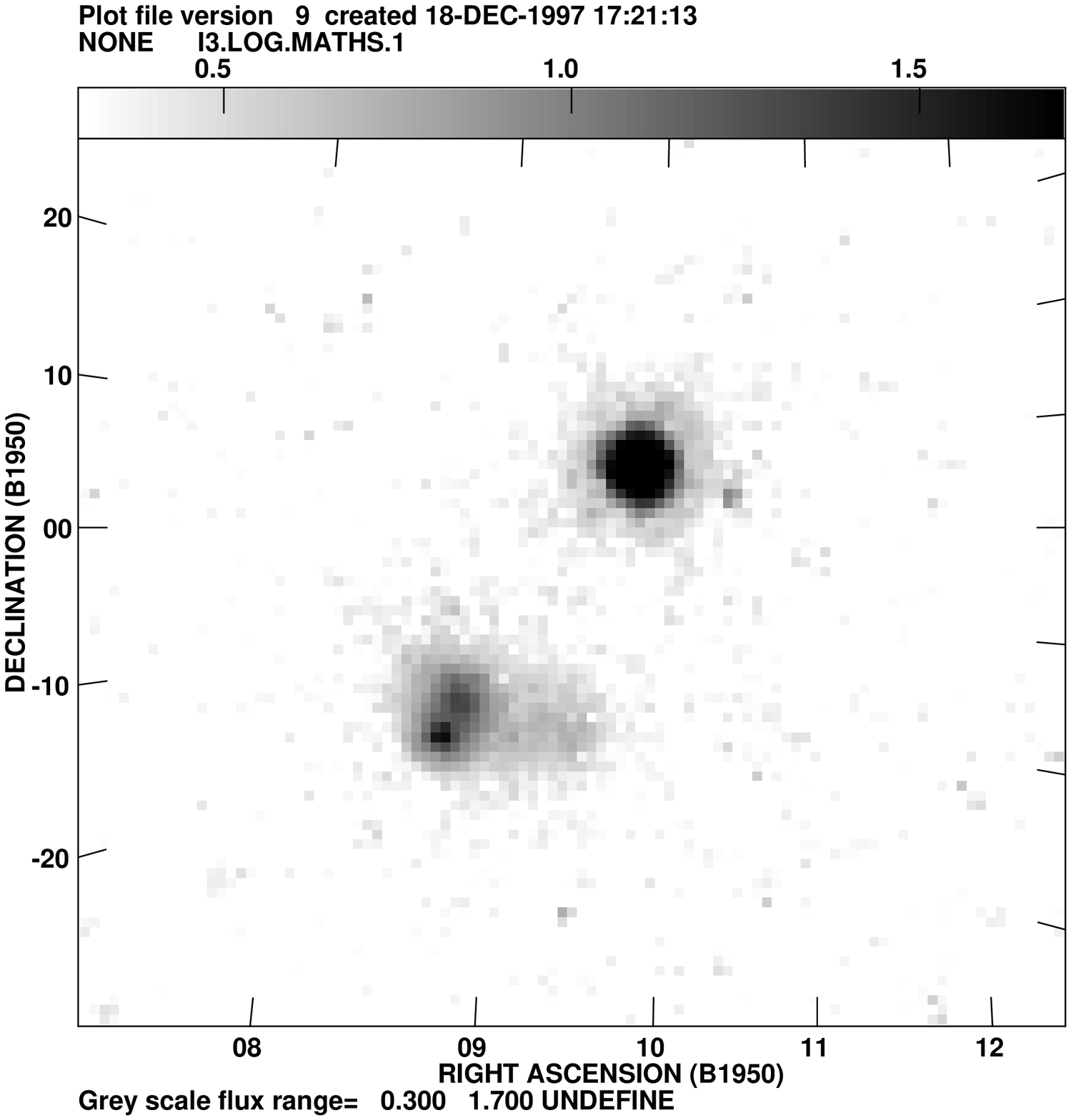}}
\put(100,100){\includegraphics{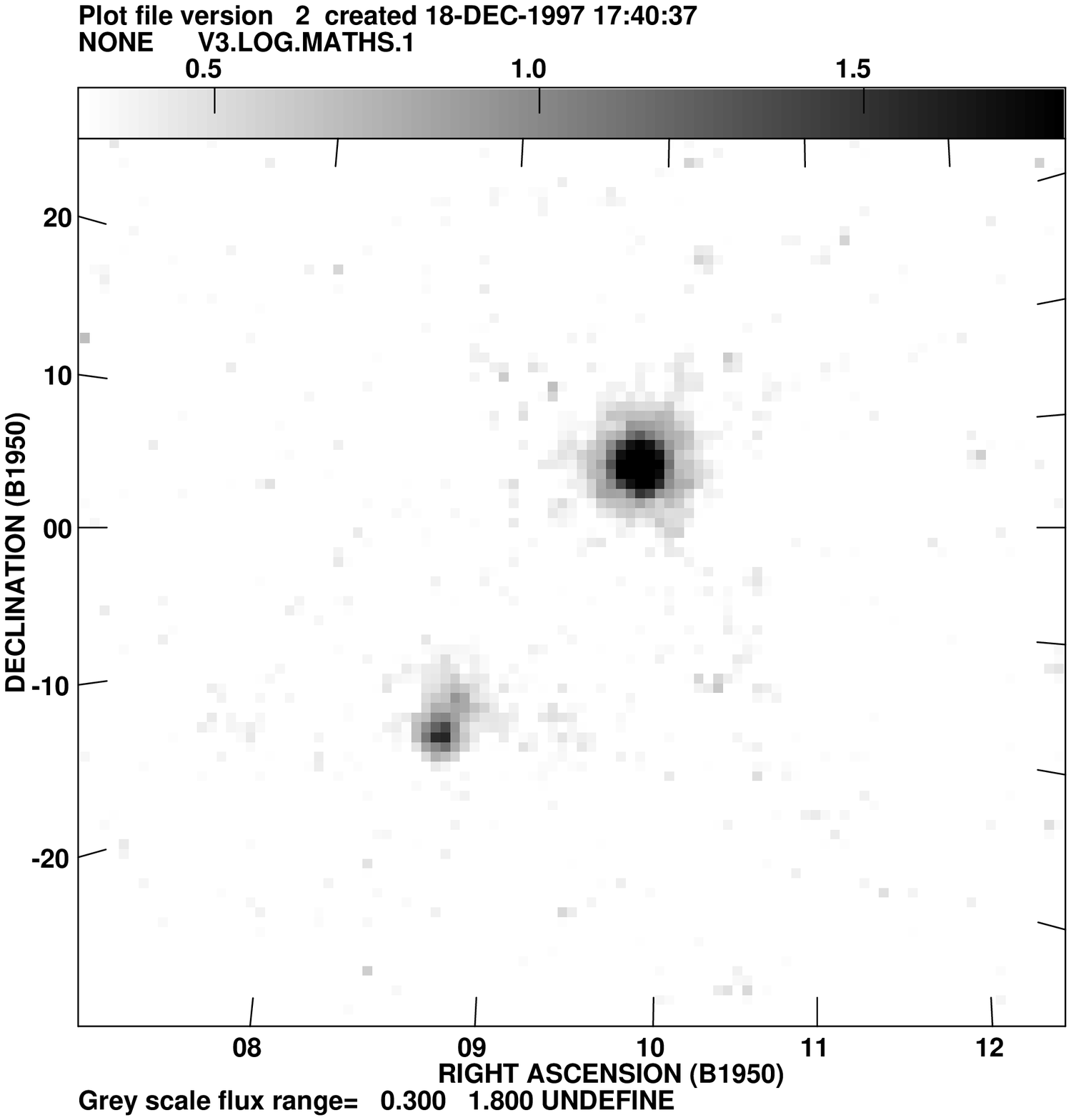}}
\end{picture}
\end{tabular}
\end{figure*}

\section{B1030+074 as a lensed system}

There is no doubt that B1030+074 
is a gravitational lens system -
the HST observations showing optical counterparts of the radio images and
the lensing galaxy itself provide incontrovertible proof.  
The flux density ratio of the two components of B1030+074
is one of the highest for any known lensed system and close to the
20:1 limit adopted for the JVAS lens search (King et al. 1998).
We attribute the different flux density ratios to the combined
effects of variability and time delay. The same explanation must apply
to the time variable flux density ratios at the same frequency. 
We plan to monitor B1030+074 to try to measure a time delay which we
will use for a determination of the Hubble constant. 

Spectra have been obtained by Fassnacht \& Cohen (1998)
using the Low Resolution Imaging Spectrograph (LRIS; Oke et al. 1995)
on the Keck telescope.  
The emission of the background source
and the lensing galaxy were spatially separated on the slit so 
spectra were extracted for each. From these Fassnacht \& Cohen (1998) find a
redshift for the lens of 0.599 and a redshift of the background source of
1.535. From the observed image splitting in the system they estimate
the mass of the lensing galaxy within its Einstein ring to be 1.3 $\times$ 
10$^{11}$/h M$_{\odot}$ (for H$_{0}$ = 100h km/s/Mpc).

We have modeled B1030+074 using a Singular Isothermal Ellipsoid
(SIE) mass distribution to describe the lens galaxy (Kormann,
Schneider and Bartelmann 1994). We place the SIE mass distribution
on the peak of the galaxy surface brightness distribution.
Since the number of free parameters (5 in total; PA and axial ratio of the surface density distribution,
velocity dispersion, and source position x,y)
is equal to the number of constraints (position x,y of image A and B, flux ratio A/B and the center 
of the galaxy; 5 in total), 
we are able to find a
mass model that reproduces the image positions and their
flux ratio (Table 1). The critical (dashed) and caustic (solid) structure
of this ``best model" are shown in Fig. 5. To investigate the stability of
this model, we performed 10,000 Monte-Carlo simulations, by adding
Gaussian distributed errors to the image positions (0.3 mas), galaxy
position (4 mas) and flux density ratio (20\%) (all 1$\sigma$ errors).
For each of the 10,000 models we solve for the mass model
parameters and source position. We also calculate the time delay
between images A and B. The resulting probability density distributions
of the lens parameters and time delay are shown in Fig. 6.
The error range on the `best model' parameters (Table 3) indicates the
range that contains 99\% of the probability density distributions
given in Fig.  6. These ranges assume that the SIE mass distribution
is the correct description of the lens galaxy.
 
We have also tried models with an extra Singular Isothermal Sphere
on the extension west of the lens galaxy, but find it has no significant
influence on the position angle of the lensing galaxy inferred
from the mass model. However, it does result in a smaller axis ratio
and velocity dispersion of the main lens mass distribution. The position
angle however appears very stable. We note that the mass
distribution is almost perpendicular to the large scale structure of the galaxy. 

The predicted time delay is around $156/h_{50}$ days with an error of only a
few percent (assuming the validity of the SIE mass model).

The mass of the lensing galaxy from the ``best mass model" inside the critical curve 
is found to be 1.552 $\times$/h 10$^{11}$ M$_{\odot}$ (for H$_{0}$ = 100h km/s/Mpc), 
consistent with the mass inside the Einstein radius quoted by Fassnacht \& Cohen (1998). 

\begin{figure*}
\begin{tabular}{cc}
\psfig{file=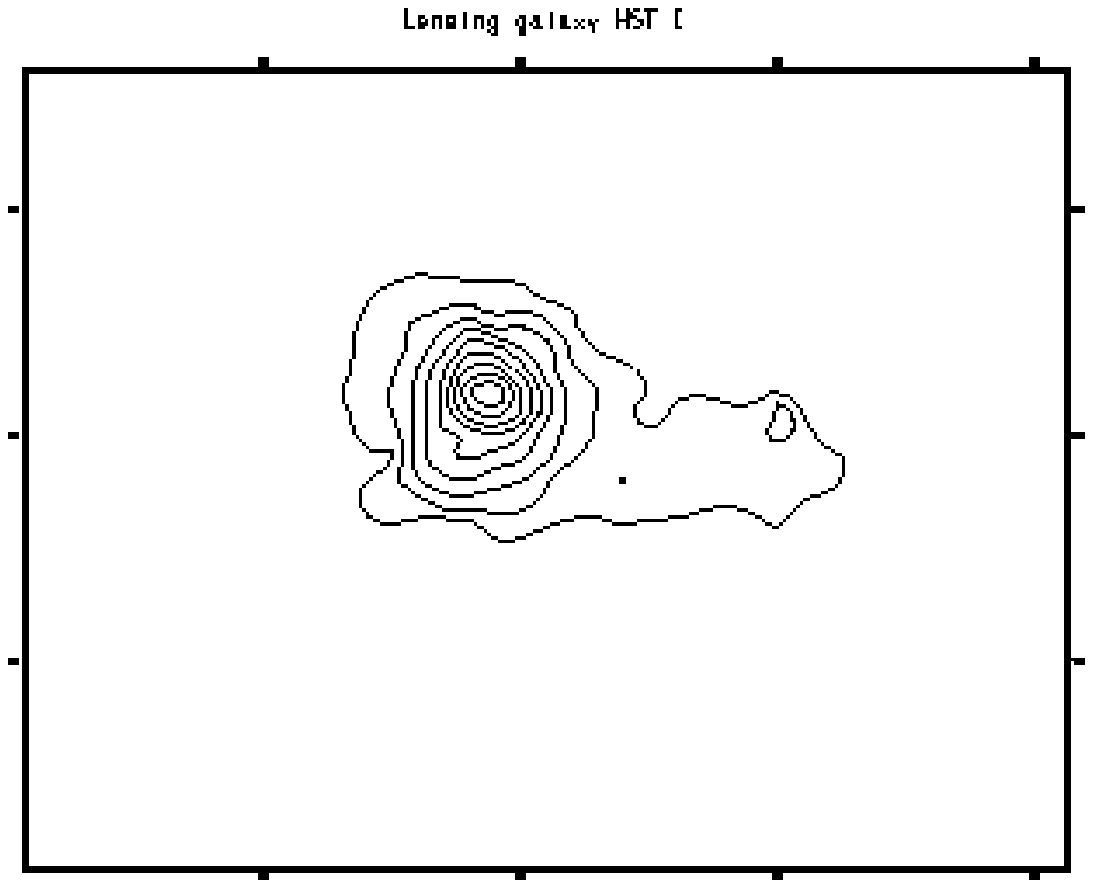,width=8.0cm,height=8.0cm,bbllx=1.8in,bblly=3.7in,bburx=7.0in,bbury=7.8in,clip=}&
\psfig{file=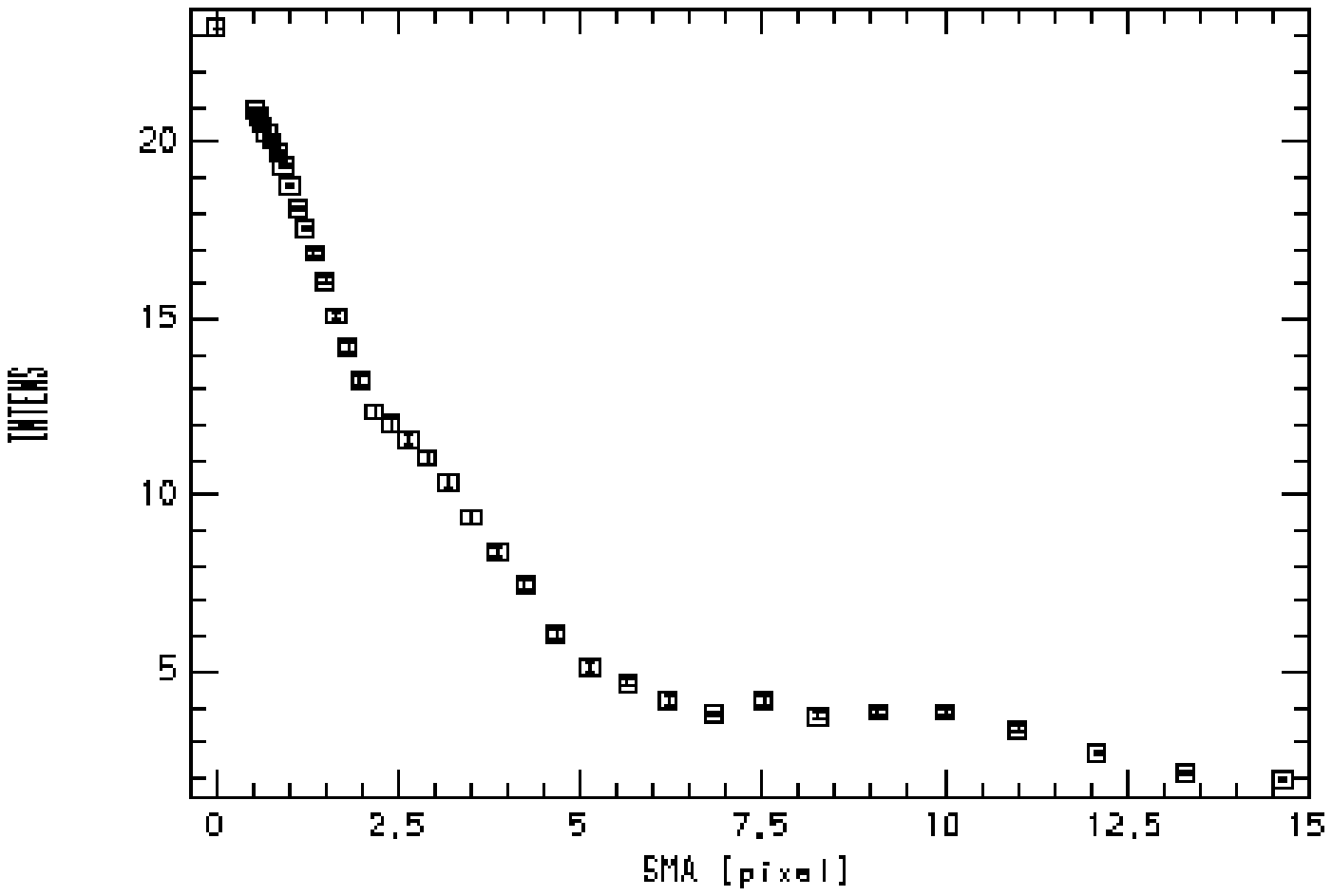,width=8.0cm,height=8.0cm,bbllx=0.9in,bblly=3.6in,bburx=7.1in,bbury=8.0in,clip=}\\
\psfig{file=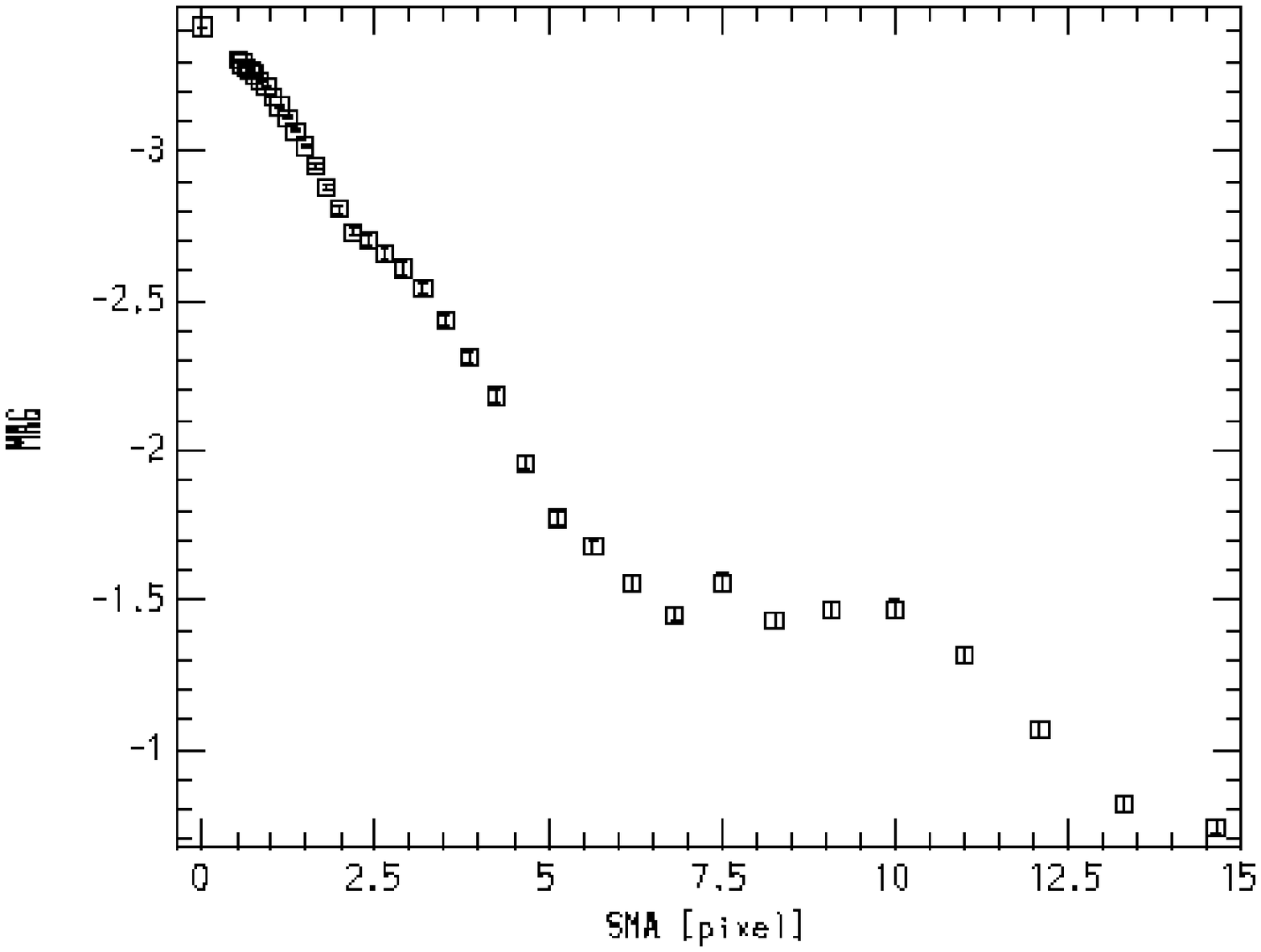,width=8.0cm,height=8.0cm,bbllx=0.42in,bblly=2.6in,bburx=7.95in,bbury=9.7in,clip=}&
\psfig{file=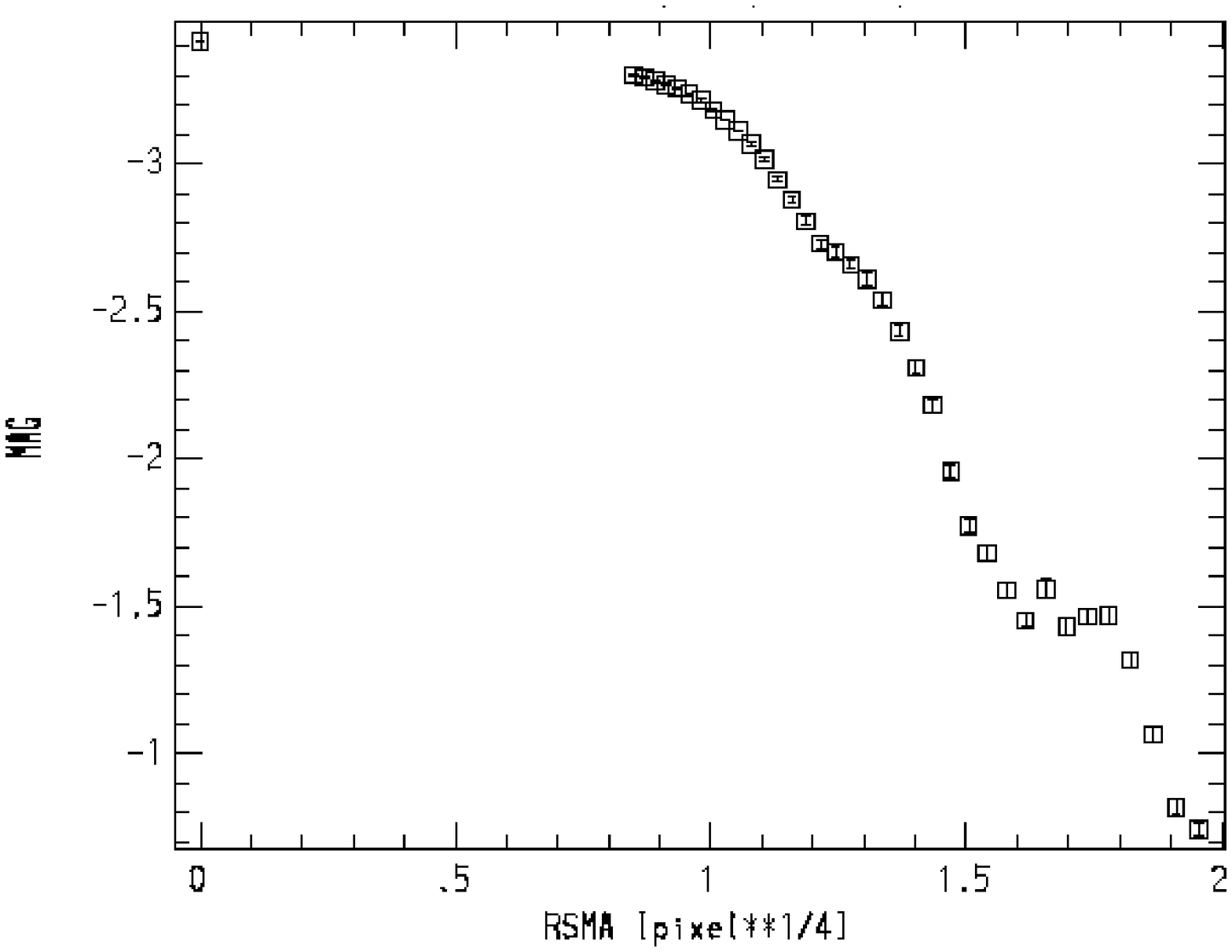,width=8.0cm,height=8.0cm,bbllx=0.7in,bblly=2.6in,bburx=7.65in,bbury=9.6in,clip=}\\
\end{tabular}
\caption{A contour plot of the lensing galaxy after the faint B component has been subtracted is
shown on the top left.
The lowest contour level is 3$\sigma$ of the sky background value and the contours differ by
a factor of 2 in intensity. The profile of the galaxy is shown on the top right. The
isophote mean
intensity as a function of the semi-major axis in pixels appears to show a hump 
due to extra light
contribution by some distorting feature in the galaxy.
At the lower left and right we have plotted the magnitude (logarithm of the isophote flux) as a
function of the semi-major axis in pixels and the the magnitude (logarithm of the isophote flux)
as a function of the semi-major axis (in pixels) to the 1/4 respectively.}
\end{figure*}

\section{Summary}
We have presented radio (VLA, MERLIN, EVN \& VLBA) results and optical
(HST) results for the new gravitational lens system B1030+074. The
radio maps all show two unresolved components except for the higher
resolution VLBA map which reveals faint jet-like structure in the
strongest component. The lensing galaxy is revealed in the HST V and I
images. It shows substructure which indicates that is it not a smooth elliptical galaxy.
B1030+074 is likely to be a lens suitable for the measurement 
of the Hubble constant. 

\begin{table*}
\caption{The ``best model" and the 99\% confidence intervals from the Monte Carlo simulations.}
\begin{tabular}{cccc}
Velocity dispersion & Surface density axis ratio & Position angle (N$\rightarrow$E)
& Time delay (B-A)\\
(km/s) &         &  (degree)   & (day/h$_{50}$)  \\
    &         &        &            \\
258.5 (+8.4) (-5.5) & 0.66 (+0.09) (-0.12) &  -10.2 (+14.7) (-10.1) & 156.1 (+2.7) (-2.5) \cr
\end{tabular}
\end{table*}

\section*{Acknowledgments}
We would like to thank Dr. Tom Muxlow for reducing the 1.7 GHz MERLIN map of B1030+074 
and Dr. Sunita Nair for constructive comments on the paper. 
This research used observations with the Hubble Space Telescope,
obtained at the Space Telescope Science Institute, which is operated
by Associated Universities for Research in Astronomy Inc. under NASA
contract NAS5-26555.  The Very Large Array is operated by Associated
Universities for Research in Astronomy Inc. on behalf of the National
Science Foundation.  
MERLIN is operated as a National Facility by NRAL,
University of Manchester, on behalf of the UK Particle Physics \&
Astronomy Research Council. 
We would also like to thank the staff of the EVN observatories and the 
MPIfR correlator for assistance with the EVN observations. 
This research was supported by European Commission, TMR Programme,
Research Network Contract ERBFMRXCT96-0034 ``CERES".

\begin{figure}
\begin{tabular}{cc}
\psfig{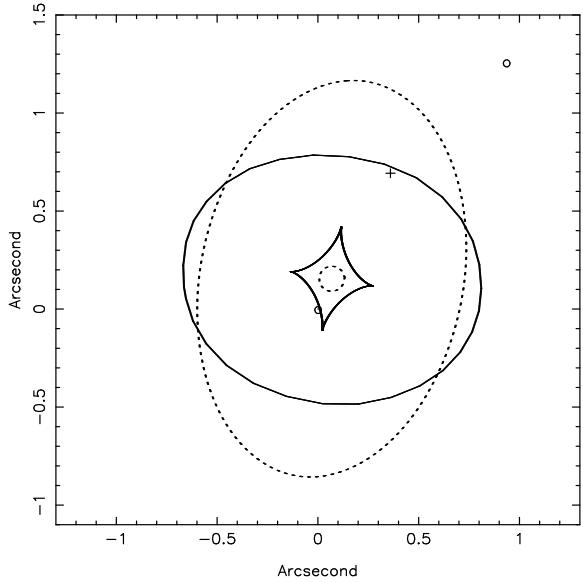}
\end{tabular}
\caption{The critical (dashed) and caustic (solid) structure of the
    ``best" model of B1030+074 (Table 3). The circles indicate the
    observed image positions, the cross the inferred source position.}
\end{figure}
 
\begin{figure}
\begin{tabular}{cc}
\psfig{file=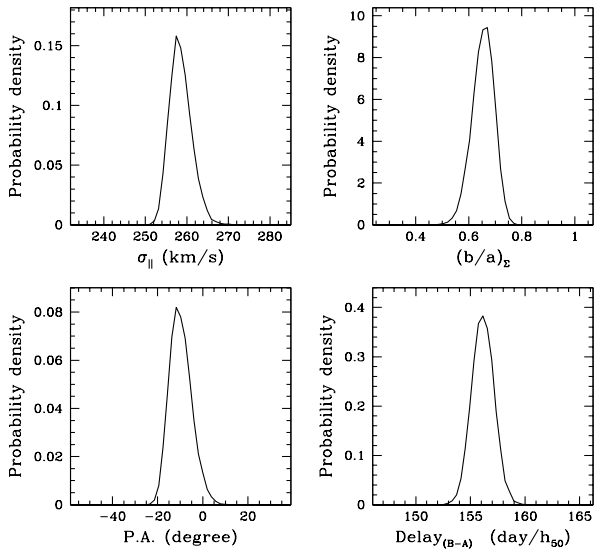,width=8cm,height=9cm,bbllx=0.3in,bblly=0.7in,bburx=8.0in,bbury=9.8in,cli
p=}
\end{tabular}
\caption{The probability density distributions of the line-of-sight velocity
    dispersion, axis ratio and position angle of the lensing galaxy,
    and the time delay between lens images A and B. These distributions
    were determined by Monte-Carlo simulations, which include all
    observational errors (see text).
    }
\end{figure}

\end{document}